\documentclass[10pt,letterpaper]{article}
\usepackage[top=0.85in,left=2.75in,footskip=0.75in]{geometry}

\usepackage{amsmath,amssymb}

\usepackage{changepage}

\usepackage{textcomp,marvosym}

\usepackage{cite}

\usepackage{nameref,hyperref}

\usepackage[right]{lineno}

\usepackage[nopatch=eqnum]{microtype}
\DisableLigatures[f]{encoding = *, family = * }

\usepackage[table]{xcolor}

\usepackage{array}

\usepackage{multirow}
\newcolumntype{+}{!{\vrule width 2pt}}

\newlength\savedwidth

\raggedright
\usepackage[aboveskip=1pt,labelfont=bf,labelsep=period,justification=raggedright,singlelinecheck=off]{caption}

\makeatletter
\renewcommand{\@biblabel}[1]{\quad#1.}
\makeatother

\usepackage{lastpage,fancyhdr,graphicx}
\usepackage{epstopdf}
\fancyheadoffset[L]{2.25in}
\fancyfootoffset[L]{2.25in}
\begin{document}
\vspace*{0.2in}

\begin{flushleft}
{\Large
\textbf\newline{Exploring language endangerment: historical, geographical, and economic insights from multilayer language-country bipartite network analysis} 
}
\newline
\\
Kazuho Nomura*,
Yuichi Ikeda
\\
\bigskip
Graduate School of Advanced Integrated Studies in Human Survivability, Kyoto University, Kyoto, Japan
\\
\bigskip

%
%





* nomura.kazuho.32r@st.kyoto-u.ac.jp

\end{flushleft}
\section*{Abstract}
Language endangerment is a phenomenon in which approximately 40\% of languages spoken worldwide are predicted to disappear within the next few decades, resulting in the loss of cultures associated with these languages. To take effective measures against language endangerment, it is essential to quantitatively understand its characteristics because it is a phenomenon in which historical, geographical, and economic factors are intricately intertwined. In this study, multilayer language-country bipartite networks are constructed using information about which countries each language is spoken in and two types of linguistic features, namely the existence of a writing system and the function within a country. In addition, percolation simulations are conducted to measure how language and country networks break down according to the extinction of languages and to identify vulnerable connections in them. In the language network of officially used languages with their writing system, the community analysis indicated that there were communities composed of languages spoken over geographically separated distances. The strength of languages revealed that the official languages in the former colonial nations, namely English, French, Spanish, Dutch, Portuguese, and Russian, still played significant roles in the formation of these communities. In the language and country networks of unofficially used languages without their writing system, the percolation simulation revealed that languages were likely to severely disappear in the Americas, and that linguistic diversity was vulnerable in affluent countries. The findings show that the analysis of multilayer language-country bipartite networks has enabled a quantitative understanding of the language endangerment occurring worldwide from historical, geographical, and economic perspectives.



\section*{Introduction}
Language extinction is defined as the state of a language that is no longer spoken, and it is a common phenomenon in human history. Languages of the ancient era, such as Sumerian, Egyptian, and Etruscan, were no longer utilized\cite{Swadesh1948}, and Hebrew, which is the official language of Israel, was eradicated by the third century\cite{Hagege2000}. Nahuatl and Quechua were utilized interchangeably with other languages in Central America, which was conquered by the Aztec Empire, and in South America, which was conquered by the Inca Empire, respectively\cite{Garza1991}. Due to the colonization of European nations, European languages are currently being disseminated throughout the globe, including English in Africa, Australia, New Zealand, and North America; French in Africa and Canada; Spanish in Central and South America; Portuguese in Brazil; and Russian in Siberia\cite{Tsunoda2006}.

Although language extinction has often occurred in history, the rate of extinction has been identified as language endangerment since the 1990s\cite{Dorian1989, Hale1992}. The extinction rate has been accelerating since the 19th century, which is considered to be associated with the development of economic globalization\cite{Ikeda2021}, and is predicted to continue even in the 21st century\cite{Simons2019}. According to the latest language data\cite{Eberhard2023a}, among the 7,010 languages spoken worldwide, 3,050 languages are classified as endangered languages, which accounts for 43.5\% of all the languages. The number of endangered language speakers is 90.2 million out of 10.8 billion people, which is less than 1\% of all. As endangered languages are no longer transmitted from their current speakers to the upcoming generation, a phenomenon known as language shift\cite{Fishman1991, Crystal2000}, it is anticipated that they will become extinct within a period of approximately one hundred years, owning to the demise of their remaining speakers. Although language endangerment is prevalent worldwide, its severity varies from one country to another. Language endangerment is particularly severe in the United States and Australia. In these countries, more than 80\% of the languages that were previously spoken in 1795 have already become extinct by the year 2018\cite{Simons2019}.

Since the 1990s, studies have been conducted on individual endangered languages, while studies on the phenomenon of language endangerment itself have only recently begun. Conventionally, endangered languages were the study topic in linguistics and anthropology, and their linguistic features, such as pronunciation, vocabulary, and grammar, were described qualitatively. As these qualitative studies of individual languages progressed, several databases were created to compile their information, for example, Ethnologue\cite{Eberhard2023a}, Glottolog\cite{Hammarstrom2023}, Automated Similarity Judgment Program (ASJP)\cite{Wichmann2022}, and World Atlas of Language Structures (WALS)\cite{Dryer2013}. The application of mathematical analysis methods to the analysis of large language data has enabled quantitative studies of the phenomenon of language endangerment itself.

In a 1992 publication\cite{Krauss1992}, a quantitative study clarified the extent of language endangerment in each region based on the percentage of endangered languages by region. It was estimated that 90\% of languages would be severely endangered or extinct by the end of the 21st century. The updated study revealed that the progress of the phenomenon varied from region to region: language endangerment is, in general, severe in the Americas, Australia, and New Zealand\cite{Simons2013}.

The regression analysis was conducted to identify variables that could explain the endangerment level of each language, and to estimate language loss in the future by incorporating demographic and environmental data\cite{Bromham2022}. This study has enabled to identify some factors that are unique to endangered languages, but they have not been thoroughly studied yet because languages become endangered for various complicated reasons. One limitation of the broad-scale model was that historical contexts such as patterns of colonial expansion were not captured as a factor of language endangerment due to the nature of the data used.

Furthermore, the network analysis was applied to networks of endangered languages, constructed according to their geographical distances\cite{Lee2022}. Given that language endangerment is primarily caused by language shift, network analysis is advantageous as it enables quantitative comprehension of the connections among languages. Although the previous study has created and analyzed networks of only endangered languages to understand their characteristics, it is necessary to cover the interconnections among endangered and non-endangered languages considering the nature of language shift.

The present study aims to quantify the characteristics of the language endangerment progressing globally based on the data of languages around the world. Multilayer bipartite networks of languages and countries are constructed based on information about which countries each language is spoken in, and two linguistic features, such as the existence of a writing system and the function within a country. In the bipartite graph, languages spoken in multiple countries create connections among geographically separated languages, which enables capturing historical and economic connections, as well as geographical connections clarified in the previous study\cite{Lee2022}. By dividing the network into multiple layers, it is feasible to conduct network analysis tailored to the specific nature of languages, such as official languages in certain countries, or languages that do not possess a writing system.

\section*{Materials and methods}
\subsection*{Data}
This study uses two types of data: Ethnologue and World Development Indicators (WDI). Ethnologue is a database for languages throughout the world, which includes information about how languages are used, who uses them, where and for what purpose\cite{Eberhard2023a}. WDI is the World Bank collection of various development indicators for countries, ranging from economics and education to environment\cite{WB2023}.

\subsubsection*{Ethnologue}
The Ethnologue data comprises three distinct categories of information, namely languages, countries, and languages within each country\cite{Eberhard2023b}. The information regarding languages encompasses its linguistic characteristics, including the language family, the existence of a writing system, the primary country and the number of countries where it is spoken, the population of its speakers, and the endangerment level for 7,615 languages. The information regarding countries includes the number of languages and the population for 242 countries or regions. In the information regarding languages within each country, languages spoken in multiple countries are counted multiple times, so 12,118 languages are listed. The information presented herein comprises the population of the speakers, the endangerment level, the existence of transmission, and the function in each country. The functions of languages delineate the official usage of each language within a country, thereby categorizing them as working languages, identity languages, languages with both functions, or recognized languages.

The Ethnologue data defines “languages” as encompassing languages that are not suitable for this study. The exclusion of a language is determined by three distinct criteria: (1) whether it still has speakers in a particular country, (2) whether it is a spoken language, not a sign language, and (3) whether it is transmitted within that country. Firstly, languages that have already become extinct are excluded, as this study focuses on the connections among existing languages at present. Secondly, sign languages are excluded since the shift from a sign language to a spoken language is impossible, and the reverse shift is impossible as well. Thirdly, this study only includes languages that are transmitted over several generations in a given country because the language shift between languages established in the country is targeted. Among the 7,615 languages included in the Ethnologue data, 7,050 were included in the analysis. Table~\ref{table1} presents a list of languages classified according to their endangerment levels, along with the number of languages classified within each label and category.

\begin{table}[!ht]
  \centering
  \caption{
  {\bf Languages categorized by endangerment level.}}
  \begin{tabular}{|c|c|c|c|c|}
    \hline
    \bf EGIDS & \multicolumn{2}{c|}{\bf 12 labels} & \multicolumn{2}{c|}{\bf 3 categories}\\ \hline
    \bf 0  & International  & 6     & \multirow{5}{*}{Institutional} & \multirow{5}{*}{487} \\ \cline{1-3}
    \bf 1  & National       & 99    & & \\ \cline{1-3}
    \bf 2  & Regional       & 44    & & \\ \cline{1-3}
    \bf 3  & Trade          & 172   & & \\ \cline{1-3}
    \bf 4  & Educational    & 166   & & \\ \hline
    \bf 5  & Written        & 1,549 & \multirow{2}{*}{Stable}        & \multirow{2}{*}{3,473} \\ \cline{1-3}
    \bf 6a & Vigorous       & 1,924 & & \\ \hline
    \bf 6b & Threatened     & 1,633 & \multirow{5}{*}{Endangered}    & \multirow{5}{*}{3,050} \\ \cline{1-3}
    \bf 7  & Shifting       & 435   & & \\ \cline{1-3}
    \bf 8a & Moribund       & 348   & & \\ \cline{1-3}
    \bf 8b & Nearly Extinct & 306   & & \\ \cline{1-3}
    \bf 9  & Dormant        & 328   & & \\ \hline
    \bf Total & \multicolumn{4}{c|}{7,010} \\
    \hline
  \end{tabular}
  \label{table1}
\end{table}

\subsubsection*{World Development Indicators}
The WDI data is available from the data bank of the World Bank\cite{WBdatabank2023}. Indicators are obtained by setting three parameters, namely Country, Series, and Time. In the Country item, there are 217 countries and regions, along with 49 aggregates, which denote the groups of countries categorized by their geographical location, economic scales, or affiliation. In the Series item, 1,477 indicators are comprised, and classified into 12 categories, namely Economic Policy \& Debt, Education, Environment, Financial Sector, Gender, Health, Infrastructure, Poverty, Private Sector \& Trade, Public Sector, Social Protection \& Labor, and Social: health. The Time item comprises 63 years, spanning from 1960 to 2022. Some data in a particular year are missing because the survey started between the 62 years, or because surveys are conducted every other year.

The definition of “countries” differs between Ethnologue and WDI, as Ethnologue comprises 242 countries, whereas WDI comprises 217 countries. Countries are divided into four groups based on the differences between their data. The first group includes countries whose names match between the two data, and 186 countries are included in this group. Another group is those where names are different, even though the same country is indicated. This study employs the names in Ethnologue for 30 countries. The third group pertains to the situation where WDI unites several countries in Ethnologue. Seven countries in Ethnologue cannot be analyzed because their data are united into other countries. The final group pertains to situations where WDI fails to record countries in Ethnologue. The 19 countries are excluded from the analysis using WDI.

\subsection*{Construction of multilayer bipartite networks}
Based on the information about which countries each language is spoken in, a bipartite graph with nodes representing languages and countries is constructed. The bipartite graph and its projected networks are layered based on two linguistic features, namely the existence of a writing system and the function within a country. Languages are classified into four layers, depending on whether they have a writing system and whether they are officially used. The division of the network into multiple layers enables a more accurate understanding of the connections among languages according to their features.

A bipartite graph consisting of $M$ countries and $N$ languages is represented by an $m \times n$ matrix. The matrix element is written as $a_{mn}=1$ when language $n$ is spoken in country $m$, and it is written as $a_{mn}=0$ when the language is not spoken in the country. The matrix $B$, which indicates the bipartite graph of language and country, is expressed in Eq(1) as follows:

\begin{equation}
    B =
    \begin{pmatrix}
        a_{11} & \hdots & a _{1n} & \hdots & a_{1N} \\
        \vdots &  & \vdots & & \vdots \\
        a_{m1} & \hdots & a_{mn} & \hdots & a_{mN} \\
        \vdots &  & \vdots & & \vdots \\
        a_{M1} & \dots & a_{Mn} & \dots & a_{MN}\\
    \end{pmatrix}.
\end{equation}

When a bipartite graph has $K$ layers, the $k$th layer is expressed as $B^{(k)}$. The sum of the matrices of $k$ layers corresponds to the original matrix, which is shown in Eq(2):

\begin{equation}
    B = \sum_{k=1}^K B^{(k)}.
\end{equation}

\noindent Here, note that the connections between different layers of the bipartite graph are not reflected in the language and country networks of each layer. Despite the discrepancy between the original language and country networks and the sum of the networks of $N$ layers, it has enabled investigating the connections among languages and countries in more detail by dividing into multiple layers.

Subsequently, from the projection of the bipartite graph, two types of networks, namely language and country networks, are created. These projected networks of the $k$th layer are represented by a biadjacency matrix, which is the product of a matrix $B^{(k)}$ for the bipartite graph, and its transposed matrix $B^{(k)T}$. The calculation of the biadjacency matrix $A^{(k)}_l$ for the language network is shown as Eq (3), and the calculation of the biadjacency matrix $A^{(k)}_c$ for the country network is shown as Eq (4):

\begin{equation}
    A^{(k)}_l = B^{(k)T} B^{(k)},
\end{equation}

\begin{equation}
    A^{(k)}_c = B^{(k)} B^{(k)T}.
\end{equation}

As this study does not consider self-loops within the networks, the diagonal elements of the matrices are corrected to zero. The matrices $A^{'(k)}_l$ and $A^{'(k)}_c$ are the biadjacency matrices without self-loops, which represent the language and country networks of the $k$th layer.

\begin{figure}[!h]
  \includegraphics{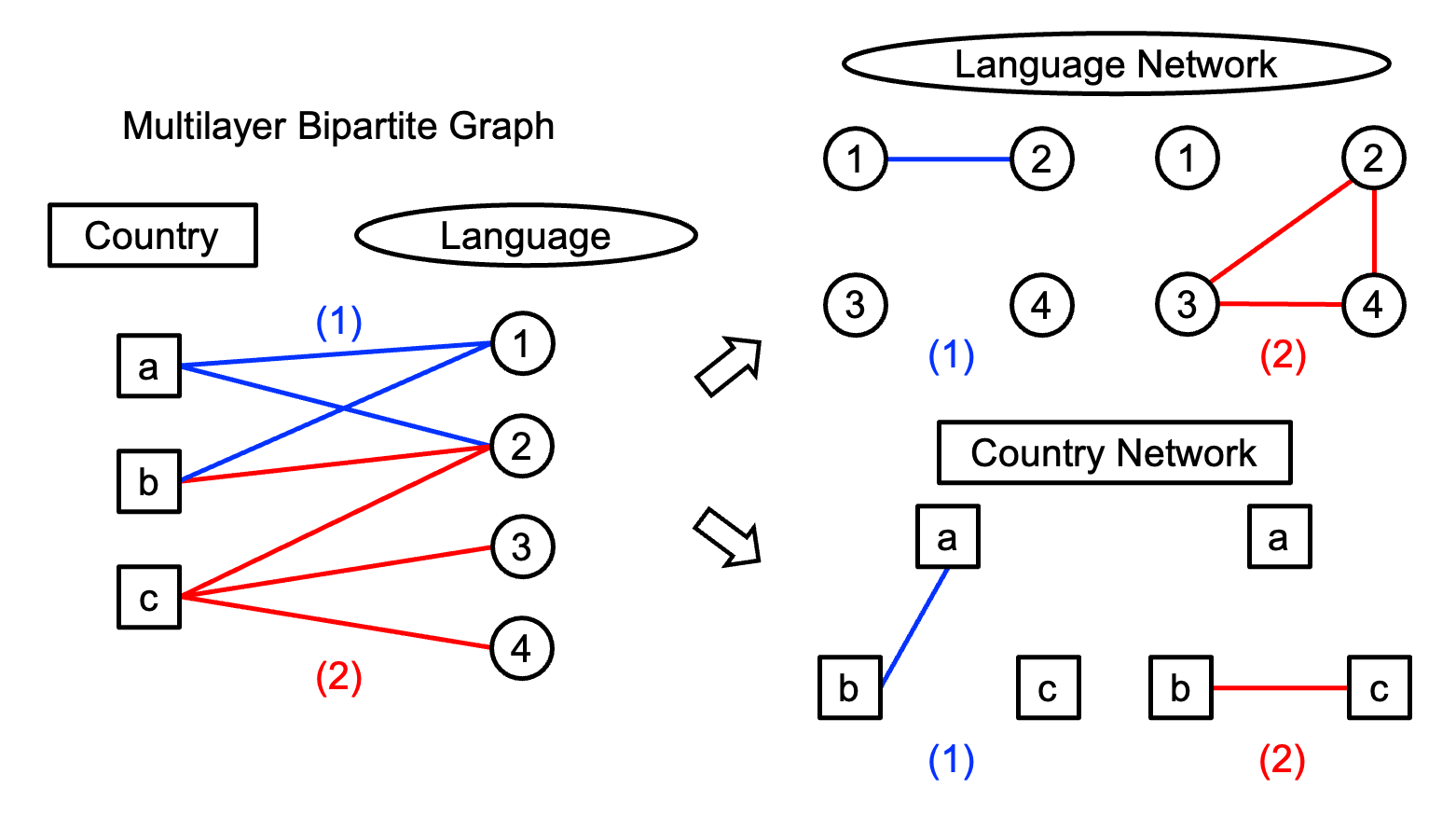}
  \caption{{\bf Construction of multilayer bipartite networks.}
  This figure comprises three components, namely the multilayer bipartite graph located on the left, the language networks located on the upper right, and the country networks located on the bottom right. The bipartite graph is constructed based on the information regarding the countries in which each language is spoken. The colors denote the multiple layers within the graph. Languages are classified into four layers based on the existence of a writing system and the function within a country. The language and country networks of each layer are constructed by projecting the bipartite graph.}
  \label{Fig1}
\end{figure}

In Fig~\ref{Fig1}, the left graph shows an example of the multilayer bipartite graph. It indicates that languages 1 and 2 are spoken in both countries $a$ and $b$, while languages 2, 3, and 4 are spoken in country $c$. The colors (1) and (2) denote multiple layers in the bipartite graph. The bipartite graph of layers (1) and (2) are described by matrices $B^{(1)}$ and $B^{(2)}$ in Eq (5) as follows:

\begin{subequations}
    \begin{equation}
        B^{(1)} =
        \begin{pmatrix}
            1 & 1 & 0 & 0 \\
            1 & 0 & 0 & 0 \\
            0 & 0 & 0 & 0
        \end{pmatrix},
    \end{equation}
    \begin{equation}
        B^{(2)} =
        \begin{pmatrix}
            0 & 0 & 0 & 0 \\
            0 & 1 & 0 & 0 \\
            0 & 1 & 1 & 1
        \end{pmatrix}.
    \end{equation}
\end{subequations}

The projected networks of language and country are represented by the product of a matrix for the bipartite graph and its transposed matrix. The calculation of biadjacency matrices $A_l$ for language networks is shown as Eq (6), and the calculation of biadjacency matrices $A_c$ for country networks is shown as Eq (7):

\begin{subequations}
    \begin{equation}
        A^{(1)}_{l} = B^{(1)T} B^{(1)} =
        \begin{pmatrix}
            1 & 1 & 0 \\
            1 & 0 & 0 \\
            0 & 0 & 0 \\
            0 & 0 & 0 
        \end{pmatrix}
        \begin{pmatrix}
            1 & 1 & 0 & 0 \\
            1 & 0 & 0 & 0 \\
            0 & 0 & 0 & 0
        \end{pmatrix} =
        \begin{pmatrix}
            2 & 1 & 0 & 0 \\
            1 & 1 & 0 & 0 \\
            0 & 0 & 0 & 0 \\
            0 & 0 & 0 & 0
        \end{pmatrix},
    \end{equation}
    \begin{equation}
        A^{(2)}_{l} = B^{(2)T} B^{(2)} =
        \begin{pmatrix}
            0 & 0 & 0 \\
            0 & 1 & 1 \\
            0 & 0 & 1 \\
            0 & 0 & 1 
        \end{pmatrix}
        \begin{pmatrix}
            0 & 0 & 0 & 0 \\
            0 & 1 & 0 & 0 \\
            0 & 1 & 1 & 1
        \end{pmatrix} =
        \begin{pmatrix}
            0 & 0 & 0 & 0 \\
            0 & 2 & 1 & 1 \\
            0 & 1 & 1 & 1 \\
            0 & 1 & 1 & 1
        \end{pmatrix},
    \end{equation}
\end{subequations}

\begin{subequations}
    \begin{equation}
        A^{(1)}_{c} = B^{(1)} B^{(1)T} =
        \begin{pmatrix}
            1 & 1 & 0 & 0 \\
            1 & 0 & 0 & 0 \\
            0 & 0 & 0 & 0
        \end{pmatrix}
        \begin{pmatrix}
            1 & 1 & 0 \\
            1 & 0 & 0 \\
            0 & 0 & 0 \\
            0 & 0 & 0 
        \end{pmatrix} =
        \begin{pmatrix}
            2 & 1 & 0 \\
            1 & 1 & 0 \\
            0 & 0 & 0
        \end{pmatrix},
    \end{equation}
    \begin{equation}
        A^{(2)}_{c} = B^{(2)} B^{(2)T} =
        \begin{pmatrix}
            0 & 0 & 0 & 0 \\
            0 & 1 & 0 & 0 \\
            0 & 1 & 1 & 1
        \end{pmatrix}
        \begin{pmatrix}
            0 & 0 & 0 \\
            0 & 1 & 1 \\
            0 & 0 & 1 \\
            0 & 0 & 1 
        \end{pmatrix} =
        \begin{pmatrix}
            0 & 0 & 0 \\
            0 & 1 & 1 \\
            0 & 1 & 3
        \end{pmatrix}.
    \end{equation}
\end{subequations}

The biadjacency matrices $A'_l$ in Eq (8) and $A'_c$ in Eq (9) are the biadjacency matrices without self-loops, which represent the links in the language and country networks shown in Fig~\ref{Fig1}:

\begin{subequations}
    \begin{equation}
        A^{'(1)}_{l} =
        \begin{pmatrix}
            0 & 1 & 0 & 0 \\
            1 & 0 & 0 & 0 \\
            0 & 0 & 0 & 0 \\
            0 & 0 & 0 & 0
        \end{pmatrix},
    \end{equation}
    \begin{equation}
        A^{'(2)}_{l} = 
        \begin{pmatrix}
            0 & 0 & 0 & 0 \\
            0 & 0 & 1 & 1 \\
            0 & 1 & 0 & 1 \\
            0 & 1 & 1 & 0
        \end{pmatrix},
    \end{equation}
\end{subequations}

\begin{subequations}
    \begin{equation}
        A^{'(1)}_{c} =
        \begin{pmatrix}
            0 & 1 & 0 \\
            1 & 0 & 0 \\
            0 & 0 & 0
        \end{pmatrix},
    \end{equation}
    \begin{equation}
        A^{'(2)}_{c} =
        \begin{pmatrix}
            0 & 0 & 0 \\
            0 & 0 & 1 \\
            0 & 1 & 0
        \end{pmatrix}.
    \end{equation}
\end{subequations}

The bipartite graph and its projected networks are layered based on two criteria derived from linguistic features. The initial criterion is whether a language possesses a writing system. Languages with their writing system are preserved in literatures, thereby enabling their transmission to the upcoming generation of their current speakers and their dissemination as second languages. In contrast, languages lacking it are typically spoken solely by the ethnic group within a restricted area. This criterion allows analyzing separately the networks of widely distributed languages and the networks of locally distributed languages.

The second criterion is what function a language has within a country. It is likely that officially used languages are transmitted at school and are learned by non-native speakers as well. However, unofficially used languages tend to be spoken only by small ethnic groups, who can speak officially used languages for their daily lives in public. As the use of a language in official settings is directly linked to its influence, this criterion facilitates the analysis of the networks of languages that exhibit strong and weak influences on other languages.

As shown in Table~\ref{table2}, there are no officially used languages without a writing system, and therefore, languages are classified into three layers. In general, Layer 1 comprises languages that are not endangered in that they are spoken in diverse circumstances and persist through their literatures. In contrast, Layer 3 comprises languages that are vulnerable to language endangerment due to their limited usage and limited geographical coverage. Layer 2 consists of different types of languages.

\begin{table}[!ht]
  \centering
  \caption{
  {\bf Three layers of languages classified by two linguistic criteria.}}
  \begin{tabular}{|c|c|c|}
    \hline
     & \bf Officially used & \bf Unofficially used\\ \hline
    \multirow{2}{*}{\bf Writing system} & Layer 1 & Layer 2\\ \cline{2-3}
     & 157 languages & 4,602 languages \\ \hline
    \multirow{2}{*}{\bf No writing system} & - & Layer 3 \\ \cline{2-3}
     & 0 languages & 2,906 languages \\ \hline
  \end{tabular}
  \label{table2}
\end{table}

In each layer, fundamental network values such as the number of nodes, links, weights, and connected components are calculated. In the case of large connected components, the average clustering coefficient and degree assortativity coefficient are calculated. Furthermore, the characteristics of connections within the network are identified through community analysis using Infomap\cite{Barabasi2019, Rosvall2008, Rosvall2009}, and strength of languages.

\subsection*{Percolation simulation of language networks}
\begin{figure}[!h]
  \includegraphics{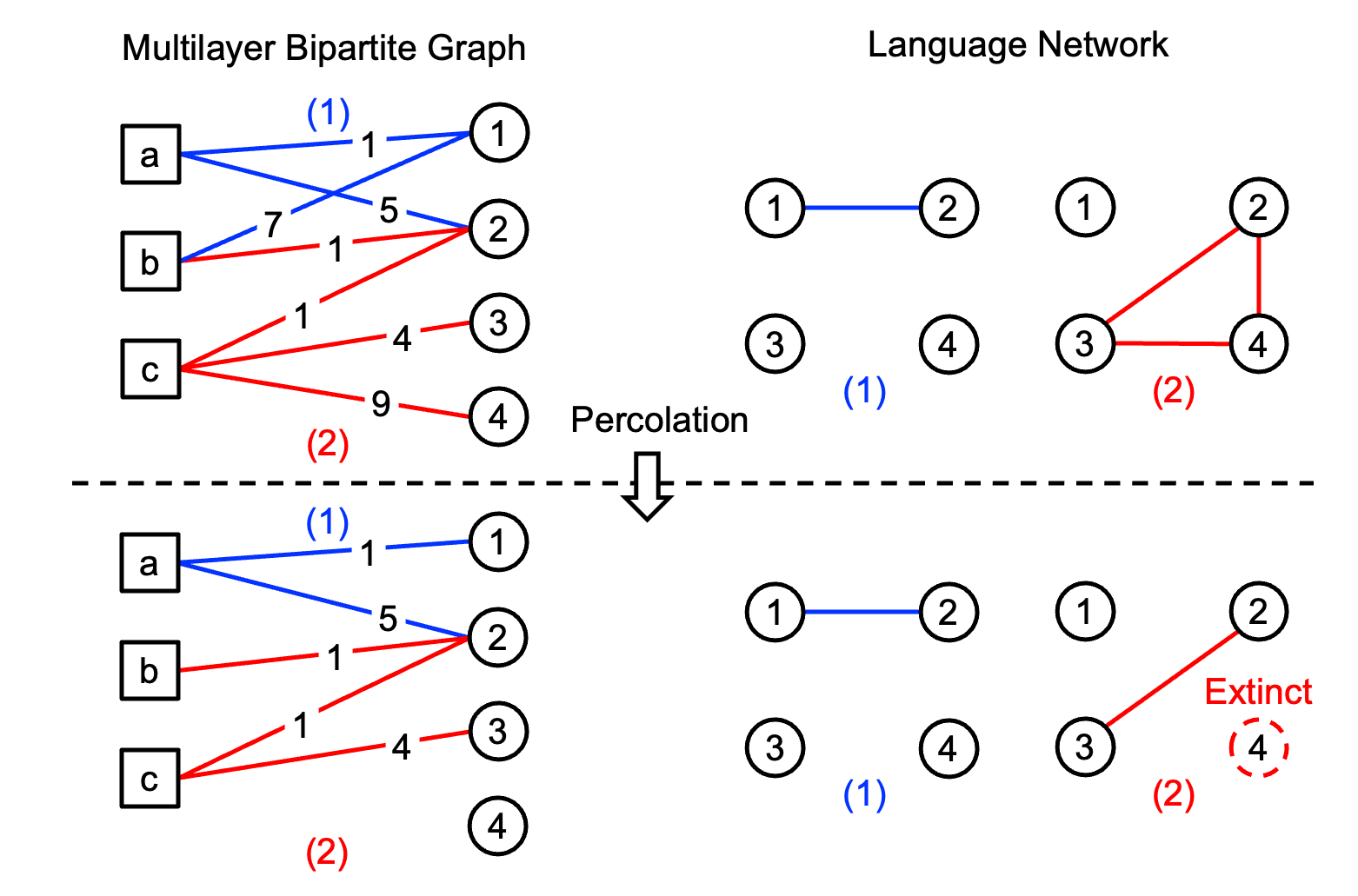}
  \caption{{\bf Percolation of links in bipartite graph.}
  This figure comprises four components. The graphs located in the upper left and upper right represent the bipartite graph and language networks before percolation. The bottom graph and networks indicate changes by percolation simulation. In the bipartite graph, links for endangered languages with endangerment levels 7 and 9 are removed. Accordingly, the links pointing towards language 4 in the language network (2) are removed. Changes in language networks indicate potentially vulnerable connections to the threat of language endangerment in the future.}
  \label{Fig2}
\end{figure}

As language endangerment progresses, certain languages spoken in certain countries are disappearing. This scenario is simulated in the bipartite graph by removing the link between a language and a country, a process known as percolation. The example of the process of percolation simulation is illustrated in Fig~\ref{Fig2}. By removing links in the bipartite graph, connections between languages are lost on the language network, which reduces the network size. Parts where the network size significantly decreases are vulnerable to language endangerment, as the connections that are currently present will be lost in the future. 

Furthermore, to comprehend the features of countries that are susceptible to language endangerment, countries are analyzed based on their GDP per capita, which indicates economic disparity, and their Gini index, which indicates economic affluence. 

\section*{Results}
\subsection*{Values of language and country networks}
\begin{figure}[!h]
  \includegraphics{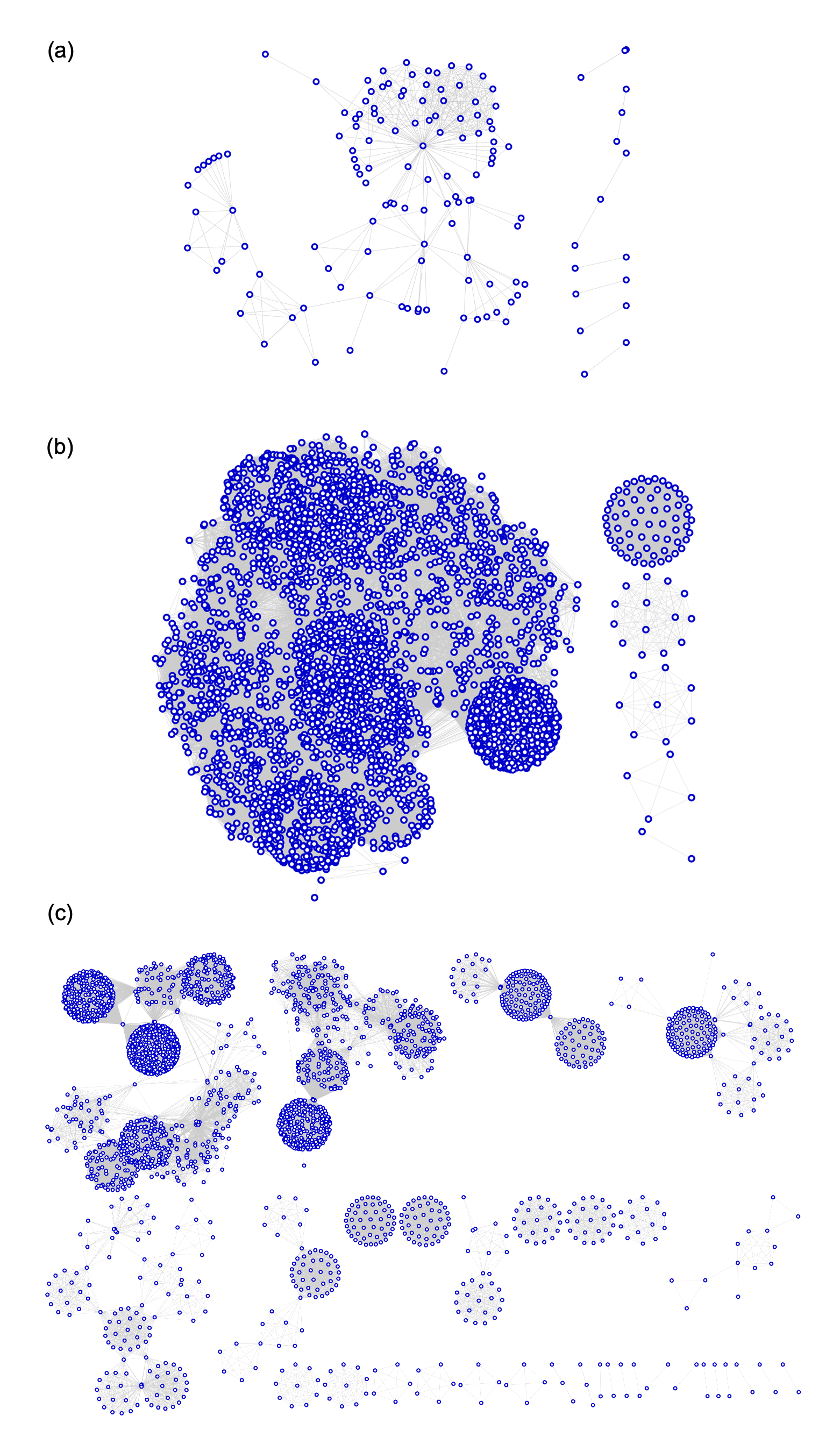}
  \caption{{\bf Language networks of three layers.}
  (a) The language network of Layer 1 was composed of 157 languages. (b) The language network of Layer 2 was composed of 4,062 languages. (c) The language network of Layer 3 was composed of 2,906 languages.}
  \label{Fig3}
\end{figure}

\begin{figure}[!h]
  \includegraphics{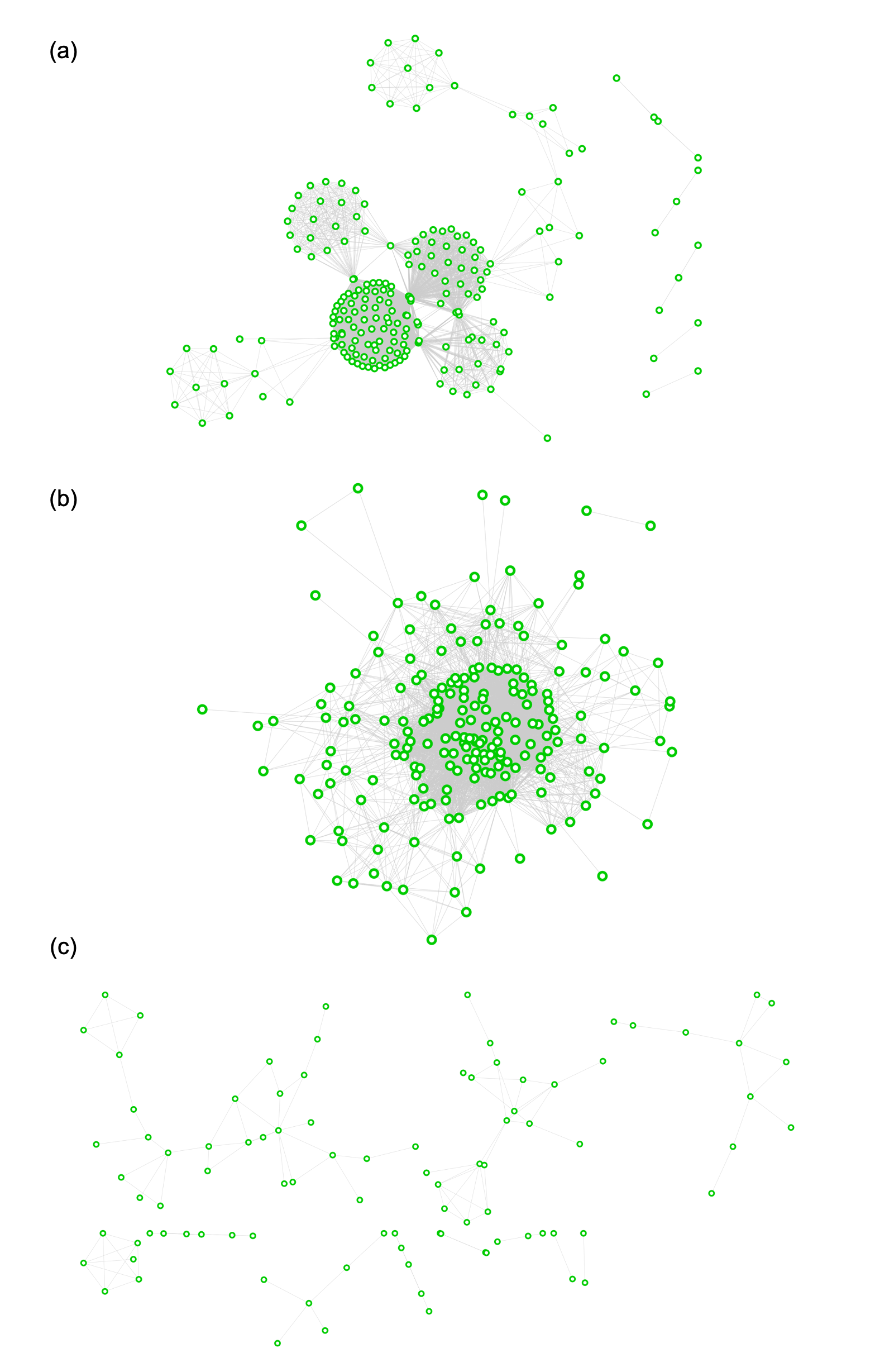}
  \caption{{\bf Country networks of three layers.}
  (a) The country network of Layer 1 was composed of 242 countries. (b) The country network of Layer 2 was composed of 218 countries. (c) The country network of Layer 3 was composed of 138 countries.}
  \label{Fig4}
\end{figure}

In Figs~\ref{Fig3} and \ref{Fig4}, the language and country networks of three layers projected from the bipartite graph are shown. The nodes are positioned in the spring-embedded layout\cite{Kamada1989}. As shown in Table~\ref{table2}, the Layer 1 network comprises officially used languages with their writing system, the Layer 2 network comprises unofficially used languages with it, and the Layer 3 network comprises unofficially used languages without it. In the case of Japan, Japanese is included in Layer 1, Ainu and Central Okinawan are included in Layer 2, and other Ryukyuan languages are included in Layer 3.

Based on the “nodes” in Table~\ref{table3}, it was discovered that Layer 1 comprised approximately 2\% of the languages, whereas Layers 2 and 3 comprised over a thousand languages. According to the “nodes” in Table~\ref{table4}, Layer 1 was a network where all countries were included, whereas Layers 2 and 3 did not include all countries. The results indicated that the unofficially used languages that made up Layers 2 and 3 were unevenly distributed only in specific countries.

\begin{table}[!ht]
  \centering
  \caption{
  {\bf Values of language networks of three layers.}}
  \begin{tabular}{|c|c|c|c|}
    \hline
    \bf Values & \bf Layer 1 & \bf Layer 2 & \bf Layer 3\\ \hline
    \bf Nodes & 157 & 4,062 & 2,906\\ \hline
    \bf Links & 374 & 391,562 & 263,173\\ \hline
    \bf Weights & 415 & 400,556 & 263,495\\ \hline
    \bf Connected components & 8 & 6 & 33\\ \hline
    \bf Isolated nodes & 19 & 8 & 23\\ \hline
    \bf Nodes in largest component & 121 & 3,962 & 1,421\\ \hline
    \bf Percentage of nodes in largest component & 77.1\% & 97.5\% & 48.9\%\\ \hline
    \bf Endangered languages & 1 & 1,166 & 1,883\\ \hline
    \bf Percentage of endangered languages & 0.637\% & 28.7\% & 64.8\%\\ \hline
    \bf Total number of speakers & $7.34 \times 10^9$ & $3.24 \times 10^9$ & $6.74 \times 10^7$\\ \hline
    \bf Average number of speakers & $1.83 \times 10^7$ & $5.84 \times 10^5$ & $2.50 \times 10^4$\\ \hline
  \end{tabular}
  \label{table3}
\end{table}

\begin{table}[!ht]
  \centering
  \caption{
  {\bf Values of country networks of three layers.}}
  \begin{tabular}{|c|c|c|c|}
    \hline
    \bf Values & \bf Layer 1 & \bf Layer 2 & \bf Layer 3\\ \hline
    \bf Nodes & 242 & 218 & 138\\ \hline
    \bf Links & 6,007 & 5,183 & 118\\ \hline
    \bf Weights & 6,081 & 8,954 & 213\\ \hline
    \bf Connected components & 6 & 2 & 11\\ \hline
    \bf Isolated nodes & 21 & 12 & 45\\ \hline
    \bf Nodes in largest component & 207 & 204 & 29\\ \hline
    \bf Percentage of nodes in largest component & 85.5\% & 93.6\% & 21.2\%\\ \hline
  \end{tabular}
  \label{table4}
\end{table}

Moreover, as indicated in Table~\ref{table3}, the “links” and “weights” of Layer 1 were approximately twice as high as those of “nodes”, whereas the “links” and “weights” of Layers 2 and 3 were approximately 10 times as high as those of nodes. In Table~\ref{table4}, it was observed that the “links” and “weights” of Layers 1 and 2 were 20 to 40 times larger than the “nodes”, whereas the “links” and “weights” of Layer 3 were approximately equivalent to the “nodes”. It was found that the Layer 2 language network had similar features to Layer 3, while the country network had similar features to Layer 1.

\begin{figure}[!h]
  \includegraphics{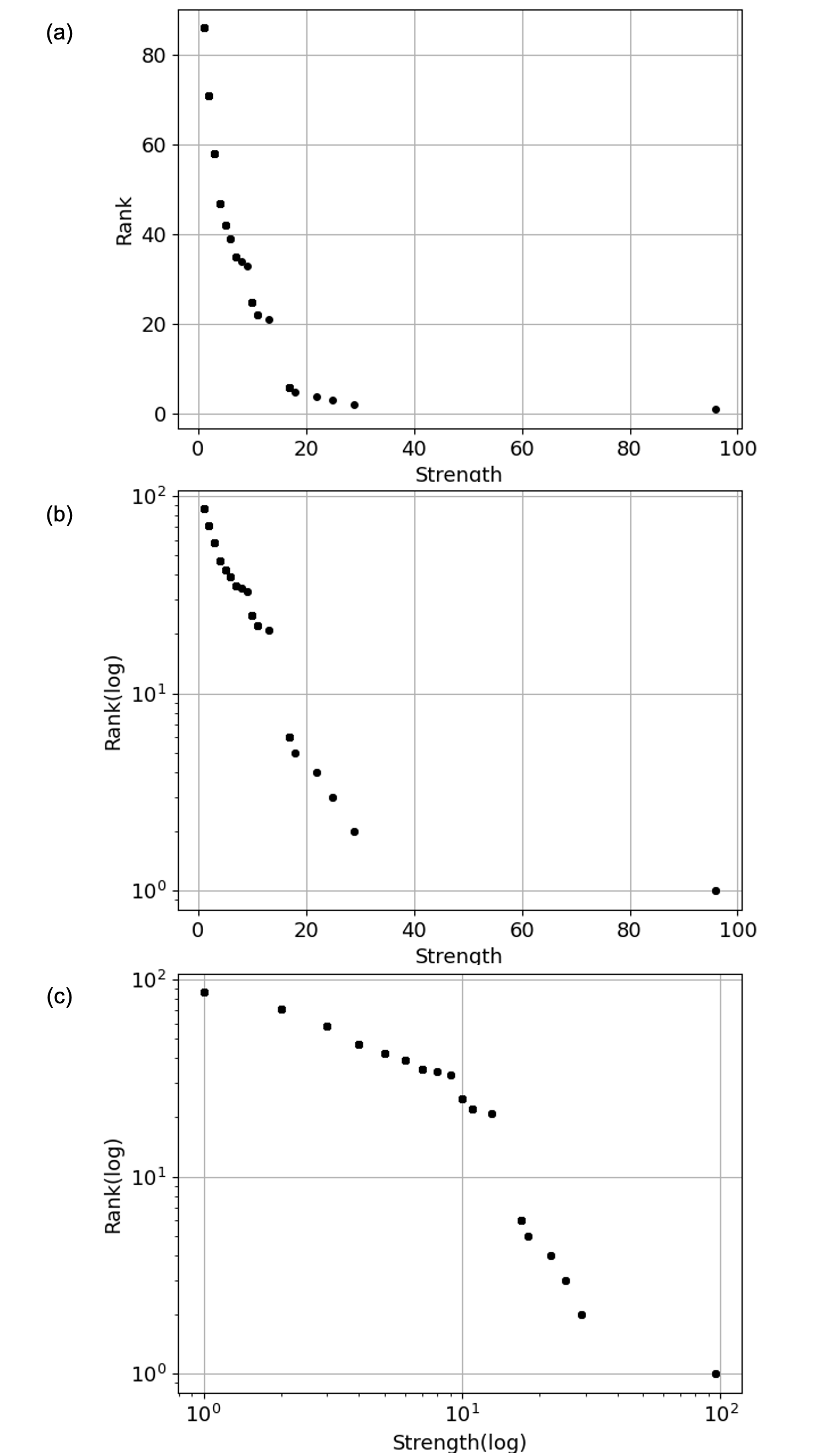}
  \caption{{\bf Rank size plot for Layer 1 language network.}
  (a) The plot is displayed linearly. (b) The plot is displayed semi-logarithmically. (c) The plot is displayed double-logarithmically. The horizontal axis represents strength linearly or logarithmically. The vertical axis represents its rank linearly or logarithmically.}
  \label{Fig5}
\end{figure}

\begin{figure}[!h]
  \includegraphics{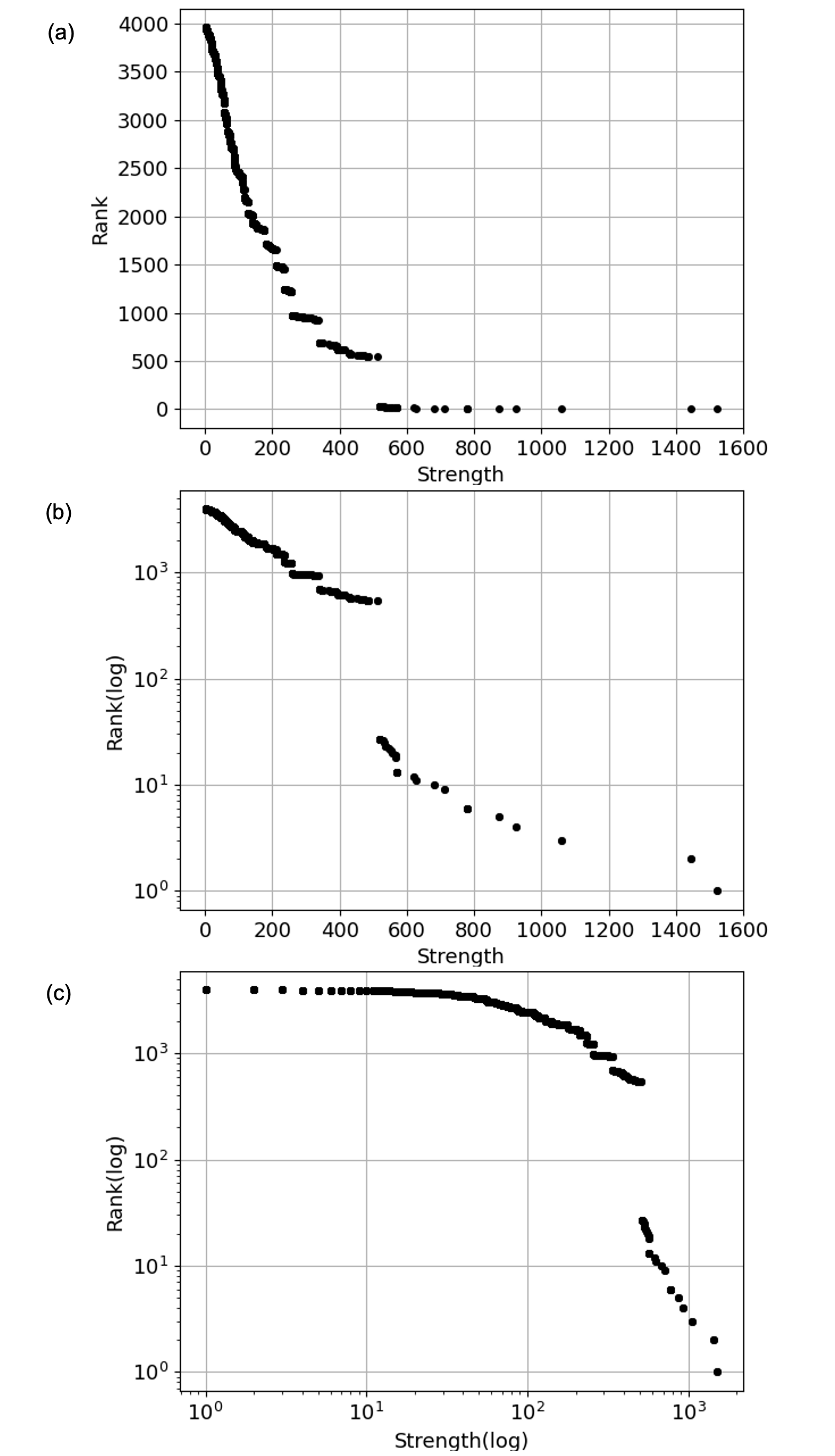}
  \caption{{\bf Rank size plot for Layer 2 language network.}
  (a) The plot is displayed linearly. (b) The plot is displayed semi-logarithmically. (c) The plot is displayed double-logarithmically. The horizontal axis represents strength linearly or logarithmically. The vertical axis represents its rank linearly or logarithmically.}
  \label{Fig6}
\end{figure}

Furthermore, from the number of “connected components” and “isolated nodes”, and “percentage of nodes in largest component”, Layer 3 was composed of many small connected components and isolated nodes. Since languages in the same country were connected in the bipartite graph, the Layer 3 network was composed of small connected components in languages and countries that were geographically close. Given that Layers 1 and 2 comprised a single large connected component as shown in Tables~\ref{table3} and \ref{table4}, and the rank size plots of strength as shown in Figs~\ref{Fig5} and \ref{Fig6} exhibited either an exponential distribution or a power distribution, it was evident that several hub languages interconnected languages and countries across the globe. According to Table~\ref{table5}, which displays the top 5 languages with the highest strength, the hubs in officially used languages in Layer 1 were English, French, Standard Arabic, Tamil, and Urdu, whereas the hubs in unofficially used languages in Layer 2 were Yue Chinese, Standard Germany, Mandarin Chinese, in addition to English and French.

\begin{table}[!ht]
  \centering
  \caption{
  {\bf High strength languages in Layers 1 and 2.}}
  \begin{tabular}{|c|c|c|c|c|c|}
    \hline
    \multicolumn{3}{|c|}{\bf Layer 1} & \multicolumn{3}{c|}{\bf Layer 2}\\ \hline
    \bf Order & \bf Language & \bf Strength & \bf Order & \bf Language & \bf Strength \\ \hline
    \bf 1 & English & 96 & \bf 1 & English & 1,544\\ \hline
    \bf 2 & French & 29 & \bf 2 & French & 1,444\\ \hline
    \bf 3 & Standard Arabic & 25 & \bf 3 & Yue Chinese & 1,059\\ \hline
    \bf 4 & Tamil & 22 & \bf 4 & Standard Germany & 924\\ \hline
    \bf 5 & Urdu & 18 & \bf 5 & Mandarin Chinese & 873\\ \hline
  \end{tabular}
  \label{table5}
\end{table}

In addition, Table~\ref{table3} provides information on the number of endangered languages and speakers. Layer 1 is composed of officially used languages, so although there are fewer constituent languages, they are non-endangered languages and on average hundreds to thousands of times more spoken than languages in other layers. English and Mandarin Chinese, especially, have billions of speakers, including native and second language speakers. On the contrary, in Layer 3, which comprises languages without a writing system, approximately 65\% of the total are endangered languages, and the number of speakers per language remains at approximately a few thousand.

Table~\ref{table6} shows two coefficients of the language networks in Layers 1 and 2: the average clustering coefficient and the degree assortativity coefficient. Positive clustering coefficients indicate a high density of the networks in Layers 1 and 2. Especially, the strongly high value of Layer 2 represents the more dense connections among languages. The degree assortativity coefficient of Layer 1 was moderately negative, whereas that of Layer 2 was positive. The negative value of Layer 1 indicates that high-degree nodes tend to establish connections with low-degree nodes, whereas the positive value of Layer 2 indicates that similar-degree nodes are linked to each other.

\begin{table}[!ht]
  \centering
  \caption{
  {\bf Coefficients in Layers 1 and 2.}}
  \begin{tabular}{|c|c|c|}
    \hline
    \bf Coefficients & \bf Layer 1 & \bf Layer 2\\ \hline
    \bf Average clustering coefficient & 0.610 & 0.926\\ \hline
    \bf Degree assortativity coefficient & -0.180 & 0.772\\ \hline
  \end{tabular}
  \label{table6}
\end{table}

\subsection*{Community analysis of language networks}
Community analysis using Infomap was conducted for the three layers of the language networks. For Layers 1 and 2, the largest connected component, which accounted for the majority of all nodes, was targeted for analysis. Regarding Layer 3, the analysis targeted up to the fifth-largest connected component, which accounted for approximately 90\% of all nodes in total. The Layer 1 language network was divided into sub-communities, and the Layers 2 and 3 language networks were divided into communities. When communities and sub-communities were classified based on the countries in which the languages were mainly spoken, communities were divided into two types: communities whose languages were locally distributed in a certain region, and communities whose languages were widely distributed across geographical distances.

The language network of Layer 1, which is composed of officially used languages with their writing system, was divided into two communities, as shown in Table~\ref{table7}. Community 2, the smaller community, was composed primarily of languages spoken in Eastern and Southern Europe and Central and Western Asia, and languages with high strength were Russian and Croatian. Community 1, on the other hand, comprised of languages spoken in the other regions, including English, French, and Arabic as languages with high strength. Each community was found to be divided into Russian-speaking and non-Russian-speaking communities based on the geographical distribution of languages.

\begin{table}[!ht]
  \centering
  \caption{
  {\bf Communities in Layer 1 language network.}}
  \begin{tabular}{|c|c|c|}
    \hline
    \bf Community & \bf 1 & \bf 2\\ \hline
    \bf Languages & 101 & 20\\ \hline
    \bf Endangered languages & 1 & 0\\ \hline
    \bf Percentage of endangered languages & 0.990\% & 0\%\\ \hline
    \bf Total number of speakers & $6.93 \times 10^9$ & $4.38 \times 10^8$\\ \hline
    \bf Average number of speakers &$6.86 \times 10^7$ & $2.19 \times 10^7$\\ \hline
    \multirow{2}{*}{\bf High-strength languages} & English, French, & \multirow{2}{*}{Russian, Croatian}\\
     & Standard Arabic & \\ \hline
  \end{tabular}
  \label{table7}
\end{table}

Community 1 in Layer 1 was divided into 13 sub-communities. There were five widely distributed sub-communities in Table~\ref{table8} and eight locally distributed sub-communities in Table~\ref{table9}. In terms of strength, it was revealed that such languages as French, Standard Arabic, English, Spanish, Dutch, Mandarin Chinese, and Portuguese were central in each sub-community. As shown in Table~\ref{table10}, Community 2 in Layer 1 was divided into two sub-communities, including a widely distributed sub-community and a locally distributed sub-community. In the widely distributed sub-community, Russian was found to be the high-strength language.

\begin{table}[!ht]
\begin{adjustwidth}{-2.25in}{0in}
  \centering
  \caption{
  {\bf Widely distributed sub-communities in Community 1 in Layer 1 language network.}}
  \begin{tabular}{|c|c|c|c|c|c|}
    \hline
    \bf Sub-community & \bf 1 & \bf 2 & \bf 5 & \bf 6 & \bf 7\\ \hline
    \bf Languages & 23 & 19 & 6 & 4 & 4\\ \hline
    \bf Endangered languages & 1 & 0 & 0 & 0 & 0\\ \hline
    \bf Percentage of endangered languages & 0.990\% & 0\% & 0\% & 0\% & 0\%\\ \hline
    \bf Total number of speakers & $1.08 \times 10^9$ & $1.58 \times 10^9$ & $5.79 \times 10^9$ & $2.64 \times 10^7$ & $1.26 \times 10^9$\\ \hline
    \bf Average number of speakers & $4.70 \times 10^7$ & $8.30 \times 10^7$ & $9.65 \times 10^7$ & $6.60 \times 10^6$ & $3.15 \times 10^8$\\ \hline
    \multirow{2}{*}{\bf High-strength languages} & French, & \multirow{2}{*}{English} & \multirow{2}{*}{Spanish} & \multirow{2}{*}{Dutch} & Mandarin Chinese,\\
     & Standard Arabic & & & & Portuguese\\ \hline
  \end{tabular}
  \label{table8}
\end{adjustwidth}
\end{table}

\begin{table}[!ht]
\begin{adjustwidth}{-2.25in}{0in}
  \centering
  \caption{
  {\bf Locally distributed sub-communities in Community 1 in Layer 1 language network.}}
  \begin{tabular}{|c|c|c|c|c|c|c|c|c|}
    \hline
    \bf Sub-community & \bf 3 & \bf 4 & \bf 8 & \bf 9 & \bf 10 & \bf 11 & \bf 12 & \bf 13 \\ \hline
    \bf Languages & 17 & 10 & 4 & 4 & 3 & 3 & 2 & 2\\ \hline
    \bf Endangered & \multirow{2}{*}{0} & \multirow{2}{*}{0} & \multirow{2}{*}{0} & \multirow{2}{*}{0} & \multirow{2}{*}{0} & \multirow{2}{*}{0} & \multirow{2}{*}{0} & \multirow{2}{*}{0}\\
    \bf languages & & & & & & & & \\ \hline 
    \bf Percentage of & \multirow{3}{*}{0\%} & \multirow{3}{*}{0\%} & \multirow{3}{*}{0\%} & \multirow{3}{*}{0\%} & \multirow{3}{*}{0\%} & \multirow{3}{*}{0\%} & \multirow{3}{*}{0\%} & \multirow{3}{*}{0\%}\\
    \bf endangered & & & & & & & & \\
    \bf languages & & & & & & & & \\ \hline
    \bf Total number & \multirow{2}{*}{$1.69 \times 10^9$} & \multirow{2}{*}{$1.26 \times 10^8$} & \multirow{2}{*}{$1.17 \times 10^5$} & \multirow{2}{*}{$1.18 \times 10^8$} & \multirow{2}{*}{$1.55 \times 10^8$} & \multirow{2}{*}{$5.10 \times 10^6$} & \multirow{2}{*}{$5.19 \times 10^5$} & \multirow{2}{*}{$8.31 \times 10^7$}\\
    \bf of speakers & & & & & & & & \\ \hline
    \bf Average number & \multirow{2}{*}{$9.96 \times 10^7$} & \multirow{2}{*}{$1.26 \times 10^7$} & \multirow{2}{*}{$2.93 \times 10^4$} & \multirow{2}{*}{$2.95 \times 10^7$} & \multirow{2}{*}{$5.18 \times 10^7$} & \multirow{2}{*}{$1.70 \times 10^6$} & \multirow{2}{*}{$2.59 \times 10^5$} & \multirow{2}{*}{$4.15 \times 10^7$}\\
    \bf of speakers & & & & & & & & \\ \hline
    \bf High-strength & \multirow{2}{*}{Tamil} & \multirow{2}{*}{Setswana} & \multirow{2}{*}{Chuukese} & Central & \multirow{2}{*}{Hausa} & \multirow{2}{*}{Mende} & \multirow{2}{*}{Welsh} & \multirow{2}{*}{Tagalog}\\
    \bf languages & & & & Kurdish & & & & \\ \hline
    \multirow{2}{*}{\bf Regions} & Southern & Southern & \multirow{2}{*}{Micronesia} & Western & Western & Western & Northern & South-Eastern\\
     & Asia & Africa & & Asia & Africa & Africa & Europe & Asia\\ \hline
  \end{tabular}
  \label{table9}
\end{adjustwidth}
\end{table}

\begin{table}[!ht]
  \centering
  \caption{
  {\bf Sub-communities in Community 2 in Layer 1 language network.}}
  \begin{tabular}{|c|c|c|}
    \hline
    \bf Sub-community & \bf 1 & \bf 2\\ \hline
    \bf Languages & 13 & 7\\ \hline
    \bf Endangered languages & 0 & 0\\ \hline
    \bf Percentage of endangered languages & 0\% & 0\%\\ \hline
    \bf Total number of speakers & $3.66 \times 10^8$ & $7.18 \times 10^7$\\ \hline
    \bf Average number of speakers &$2.82 \times 10^7$ & $1.06 \times 10^7$\\ \hline
    \bf High-strength languages & Russian & Croatian\\ \hline
    \bf Distribution & Wide & Local\\ \hline
    \bf Regions & - & Eastern and Southern Europe\\ \hline
  \end{tabular}
  \label{table10}
\end{table}

Regarding the sub-communities of the Layer 1 language network, it was discovered that six sub-communities in total were composed of languages that were widely spoken. Six languages held significant positions within these sub-communities, namely English, French, Spanish, Dutch, Portuguese, and Russian. Since these languages were the languages of the countries that were suzerain during the colonial period, it can be inferred that the widely distributed sub-communities had historical connections.

The Layer 2 language network, which is composed of unofficially used languages with their writing system, was divided into five communities, as shown in Table~\ref{table11}. As shown in Table~\ref{table6}, the average clustering coefficient of the Layer 2 language network was approximately one, and the number of constituent languages is 26 times higher than that of the Layer 1 language network, rendering it unfeasible to classify communities with the same clarity as that of Layer 1. The languages spoken in other regions than those listed in Table~\ref{table11} were also included in each community, but the majority of them were found in these regions. As in communities 2 and 4 and communities 3 and 5, languages spoken in the same region were classified as constituent languages of separate communities, making it questionable whether there were community structures in Layer 2. Nonetheless, community 1 encompasses languages from diverse regions, and these languages were connected by high-strength languages such as English, French, and Standard Germany, which were spoken in numerous countries, comprising 170, 102, and 46 countries, respectively.

\begin{table}[!ht]
\begin{adjustwidth}{-2.25in}{0in}
  \centering
  \caption{
  {\bf Communities in Layer 2 language network.}}
  \begin{tabular}{|c|c|c|c|c|c|}
    \hline
    \bf Community & \bf 1 & \bf 2 & \bf 3 & \bf 4 & \bf 5\\ \hline
    \bf Languages & 1,070 & 956 & 720 & 648 & 568\\ \hline
    \bf Endangered languages & 523 & 292 & 78 & 172 & 62\\ \hline
    \bf Percentage of & \multirow{2}{*}{48.9\%} & \multirow{2}{*}{30.5\%} & \multirow{2}{*}{10.8\%} & \multirow{2}{*}{26.5\%} & \multirow{2}{*}{10.9\%}\\ 
    \bf endangered languages & & & & &\\ \hline
    \bf Total number & \multirow{2}{*}{$4.14 \times 10^9$} & \multirow{2}{*}{$1.98 \times 10^9$} & \multirow{2}{*}{$5.66 \times 10^8$} & \multirow{2}{*}{$2.63 \times 10^9$} & \multirow{2}{*}{$1.08 \times 10^9$}\\
    \bf of speakers & & & & & \\ \hline
    \bf Average number & \multirow{2}{*}{$3.87 \times 10^6$} & \multirow{2}{*}{$2.07 \times 10^6$} & \multirow{2}{*}{$7.86 \times 10^5$} & \multirow{2}{*}{$4.05 \times 10^6$} & \multirow{2}{*}{$1.90 \times 10^6$}\\
    \bf of speakers & & & & & \\ \hline
    \bf High-strength & English, French, & Yue Chinese, & Adamawa & \multirow{2}{*}{Hakka Chinese} & Standard\\
     \bf lanugages & Standard Germany & Mandarin Chinese, & Fulfulde & & Arabic\\ \hline
     \multirow{3}{*}{\bf Main regions} & Americas, Eastern & Eastern, Southern, & North, Middle, & Eastern, Southern, & Eastern and\\
      & Europe, Australia & South-Eastern & and Western & South-Eastern, & Middle Africa,\\
      & \& New Zealand & Asia, Polynesia & Africa & Asia, Polynesia & Melanesia\\ \hline
  \end{tabular}
  \label{table11}
\end{adjustwidth}
\end{table}

In the Layer 3 language network, which is composed of unofficially used languages without their writing system, up to the fifth-largest network was targeted for community analysis, as shown in Table~\ref{table12}. Languages without their writing system were, by their nature, used only in specific regions and were not spoken in multiple geographically separated regions. Consequently, each network consisted of only languages spoken in a specific region. In the third and subsequent networks, which were relatively small, one community and one country generally corresponded to each other.

\begin{table}[!ht]
\begin{adjustwidth}{-2.25in}{0in}
  \centering
  \caption{
  {\bf Five language networks in Layer 3.}}
  \begin{tabular}{|c|c|c|c|c|c|}
    \hline
    \bf Network & \bf Layer 3-1 & \bf Layer 3-2 & \bf Layer 3-3 & \bf Layer 3-4 & \bf Layer 3-5\\ \hline
    \bf Languages & 1,421 & 688 & 182 & 148 & 124\\ \hline
    \bf Endangered languages & 980 & 314 & 178 & 125 & 41\\ \hline
    \bf Percentage of & \multirow{2}{*}{69.0\%} & \multirow{2}{*}{45.6\%} & \multirow{2}{*}{97.8\%} & \multirow{2}{*}{84.5\%} & \multirow{2}{*}{32.3\%}\\
    \bf endangered languages & & & & & \\ \hline
    \bf Total number of speakers & $2.88 \times 10^7$ & $1.39 \times 10^7$ & $1.08 \times 10^5$ & $2.66 \times 10^5$ & $1.12 \times 10^6$\\ \hline
    \bf Average number of speakers & $2.03 \times 10^4$ & $2.02 \times 10^4$ & $5.96 \times 10^2$ & $1.80 \times 10^3$ & $9.00 \times 10^4$\\ \hline
    \bf High-strength languages & Ninggerum & Bata & Central Ojibwa & Tariana & Shempire S\'{e}noufo\\ \hline
    \multirow{2}{*}{\bf Regions} & \multirow{2}{*}{Asia, Pacific} & Middle and & Northern and & \multirow{2}{*}{South America} & \multirow{2}{*}{Western Africa}\\
     & & Southern Africa & Central America & & \\ \hline
     \bf Communities & 2 & 2 & 3 & 5 & 8\\ \hline
  \end{tabular}
  \label{table12}
\end{adjustwidth}
\end{table}

\subsection*{Percolation simulation of language networks}
\begin{figure}[!h]
  \includegraphics{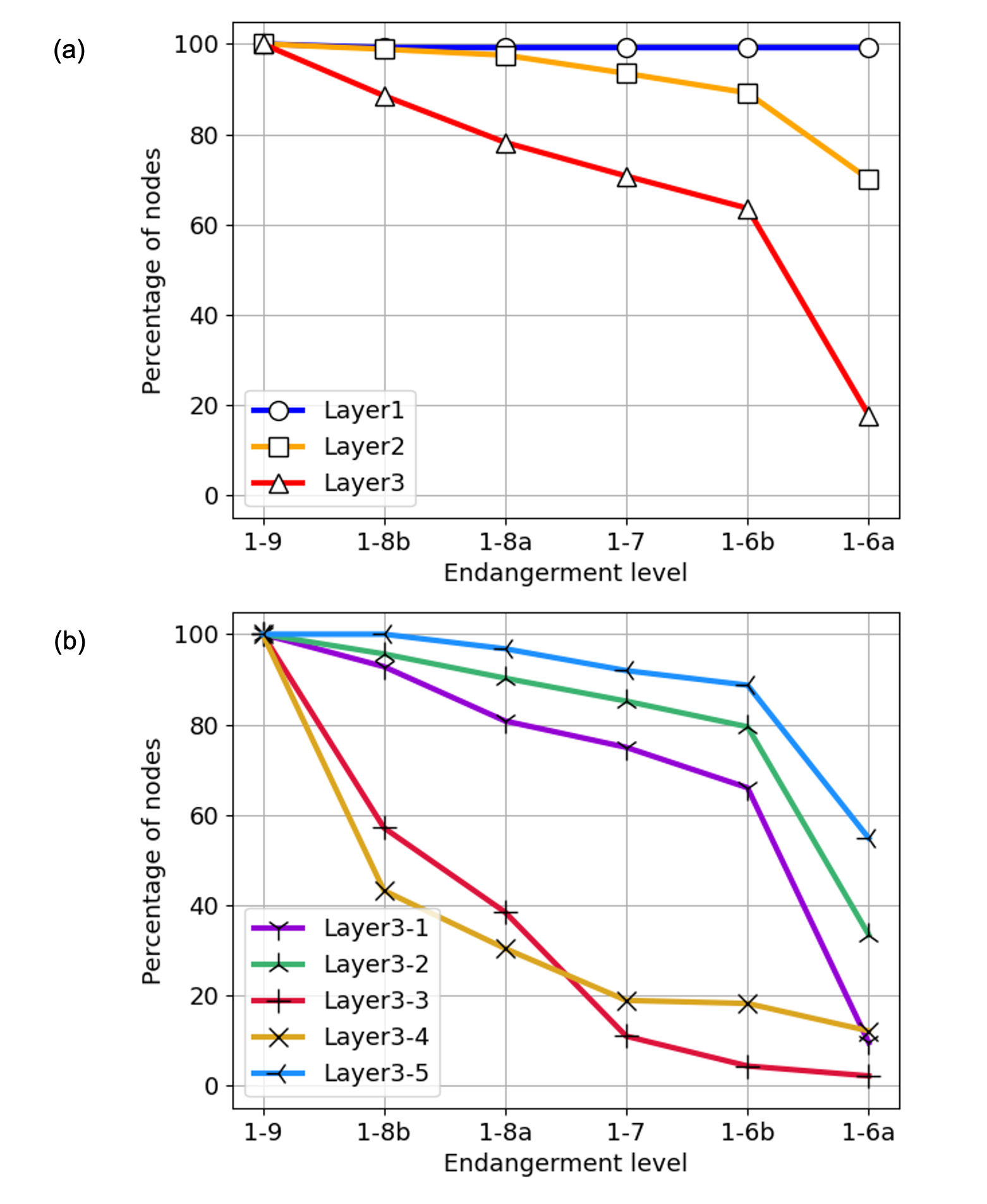}
  \caption{{\bf Change rate of nodes in language networks by percolation.}
  (a) Change rate of nodes in three language networks is indicated. (b) Change rate of nodes in five language networks of Layer 3 is indicated. The horizontal axis indicates the endangerment level for the languages included in the percolation simulation. The left column contains languages of all endangerment levels, and languages of higher endangerment levels are removed while moving to the right. The vertical axis indicates the network size. The initial network size is standardized at 100, and it changes when languages of a particular endangerment level are eliminated.}
  \label{Fig7}
\end{figure}

The future threat of language endangerment was simulated through percolation in language networks. The results of percolation in the language networks of three layers and the five language networks in Layer 3 in Fig~\ref{Fig7}. The remarkable collapse was observed in the language network of Layer 3, which had the highest proportion of endangered languages among the three layers. There were two types of size reduction within the five language networks of Layer 3. The notable reduction was observed in the language networks of Layer 3-3 and 3-4, which were composed of languages spoken in the Americas. As indicated in Fig~\ref{Fig7}, networks composed of languages that are unofficially used without their writing system will be strongly affected by the future language endangerment. Especially, it was found that the influence of the network of languages spoken in the Americas was the strongest among these.

Moreover, the percolation simulation assessed the economic characteristics of the countries where the languages removed through percolation were spoken. The countries in which the languages included in the five language networks of Layer 3 were spoken were analyzed using scatter plots, with Gini index and GDP per capita as the axes. Through the process of percolation, language nodes are removed from language networks based on their endangerment levels. The country plots disappear from the scatter plots once all languages spoken in the country have been removed.

\begin{figure}[!h]
  \includegraphics{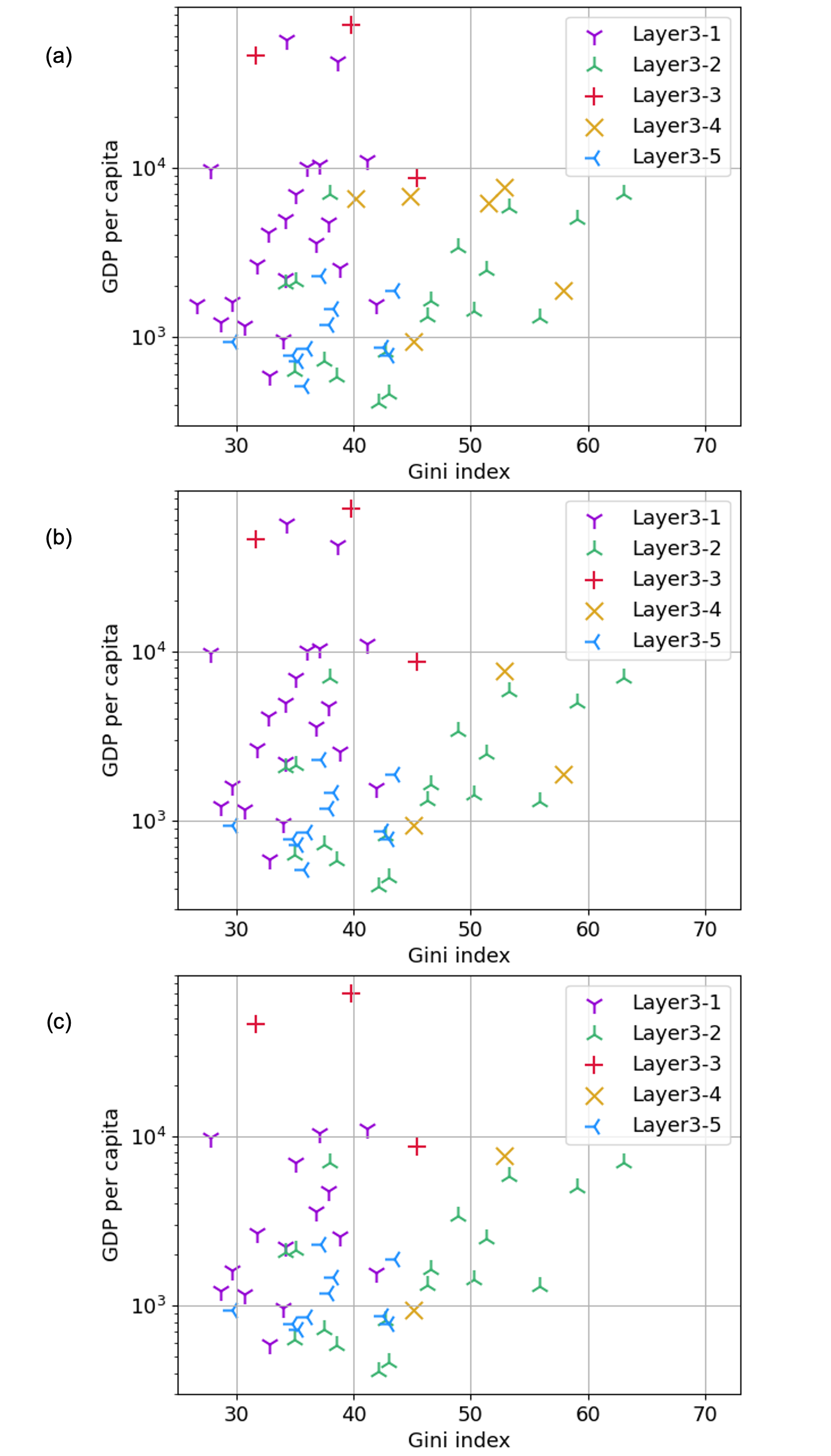}
  \caption{{\bf Country plots of languages in Layer 3 networks.}
  (a) No languages are removed. (b) Languages of level 9 are removed. (c) Languages of Levels 8b and 9 are removed. Gini index and GDP per capita of countries where languages included in language networks are spoken on the horizontal and vertical axes. The plot is removed when all languages in a country are removed through percolation.}
  \label{Fig8}
\end{figure}

\begin{figure}[!h]
  \includegraphics{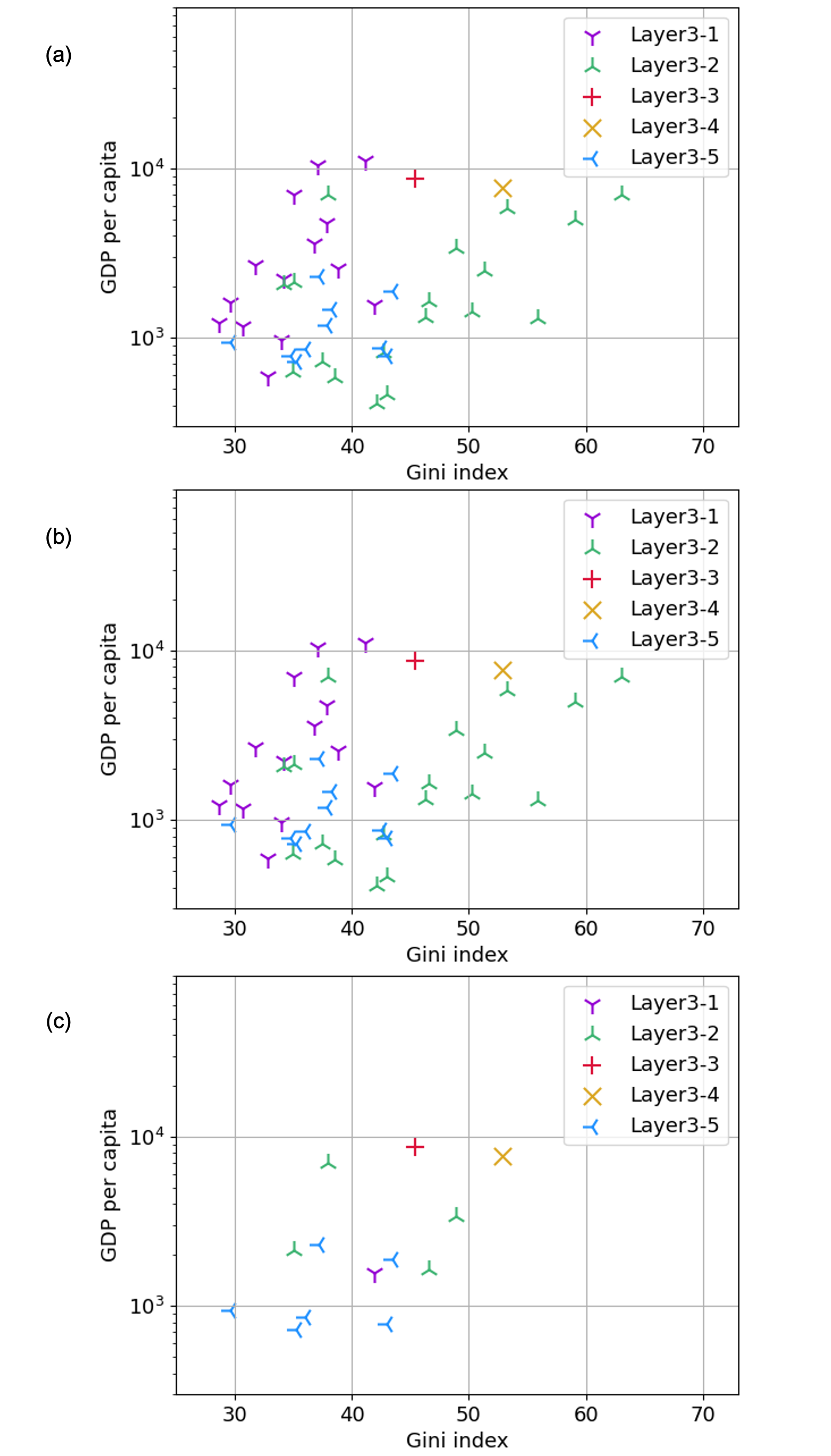}
  \caption{{\bf Country plots of languages in Layer 3 networks.}
  (a) Languages of levels 8a to 9 are removed. (b) Languages of levels 7 to 9 are removed. (c) Languages of Levels 6b to 9 are removed. Gini index and GDP per capita of countries where languages included in language networks are spoken on the horizontal and vertical axes. The plot is removed when all languages in a country are removed through percolation.}
  \label{Fig9}
\end{figure}

The changes in scatter plots of the countries where languages included in the Layer 3 language networks are spoken through the percolation simulation are shown in Figs~\ref{Fig8} and \ref{Fig9}. As highlighted by comparing the results before percolation and when languages up to endangerment levels 7 have been removed, plots of countries with high GDP per capita disappeared, regardless of Gini index. Countries whose plots were deleted included the United States, Canada, Australia, and Israel. This suggests that the potential impact of language endangerment in the future in affluent countries will be significant, regardless of economic disparities.

\section*{Discussion}
Based on the results, the characteristics of language endangerment are examined from three perspectives: historical, geographical, and economic. Historically, it is pointed out that the colonialism of European countries continues to have an impact on the officially used languages of the countries they formerly dominated. Geographically, the effects of nationalism in Europe and geographical isolation in island countries, as well as dual dominations in the Americas, are suggested. Economically, it is noted that language shifts have been brought about by the movement of people from rural areas to cities in affluent countries. Below, three characteristics of language endangerment are discussed.

Firstly, the historical characteristic of language endangerment is pointed out based on the result of community analysis of the Layer 1 language network. The primary cause of language endangerment is the language shift from the original language of the speakers to a more influential language. Given that officially used languages are likely to be the target of language shift, the characteristics of the Layer 1 language network represent the connections among influential languages. Within the sub-communities composed of widely distributed languages in the Layer 1 language network, French, Standard Arabic, English, Spanish, Dutch, Mandarin Chinese, Portuguese, and Russian were found to be central languages in terms of strength. The European languages English, French, Spanish, Portuguese, Dutch, and Russian are spoken in numerous countries worldwide in addition to their original countries, United Kingdom, France, Spain, Portugal, the Netherlands, and Russia. Countries where these languages are spoken were previously governed by these European suzerains, and even after independence, the languages of the suzerains retain their official status.

Particularly, the majority of countries situated on the African continent have experienced colonial dominance, resulting in a multilingual situation where individuals use both the official language of their suzerain and the indigenous language for daily life\cite{Calvet2010}. Each country on the African continent is composed of diverse ethnic groups, and it is unclear which ethnic language should be designated as the official language. The difficulties in African countries regarding official languages stem from the inadequate national consciousness of the diverse ethnic groups that coexist within national boundaries that have been established without regard to their local history and culture\cite{Matsumoto2016}. Although political colonialism has ended, linguistic colonialism still exists.

Secondly, the geographical characteristic of language endangerment is explained based on the geographical distribution and the result of percolation simulation of the Layer 3 language network. Table~\ref{table12} provided an overview of the geographical distribution of languages that were included in the Layer 3 language network. The languages in this layer lack a writing system, and therefore it is unlikely that they are propagated through books and used by ethnic groups apart from the original one. Language networks were created in Asia, Africa, the Americas, and the Pacific, but not in European countries and island countries, such as Japan. This result implies that linguistic diversity persists in Asia, Africa, the Americas, and the Pacific, whereas it has been diminishing in Europe and island countries. Linguistic diversity is maintained in certain regions due to the presence of numerous languages, including 708 languages in Indonesia, 451 in India, and 304 in China in Asia, 528 languages in Nigeria, 278 in Cameroon, and 215 in the Democratic Republic of the Congo in Africa, 291 languages in Mexico, 228 in the United States, and 217 in Brazil in the Americas, 839 languages in Papua New Guinea and 219 in Australia in the Pacific.

The loss of linguistic diversity in Europe and Japan has been considered to be due to nationalism, which eliminated various languages to promote the penetration of national languages with its slogan of “forming a people through their language, and forming a nation through its people”\cite{Matsumoto2016}. In the past, the spread of a particular language has restricted the usage of others\cite{Kaplan2008}, as in the prohibition of Ryukyuan languages by “dialect tags” in Okinawa, Japan\cite{Mashiko2014}.

Furthermore, as shown in Fig~\ref{Fig7}, the percolation simulation revealed that the size of the language networks in the Americas is rapidly decreasing, which implies that these regions are likely to be greatly affected by language endangerment. Before European colonial rule began in the 15th century, the Americas were ruled by the Inca and Aztec empires\cite{Garza1991}. This dual dominance is considered to make the Americas the most vulnerable to language endangerment in the future.

Thirdly, the economic characteristic of language endangerment is suggested based on the result of percolation simulation of countries where languages included in the Layer 3 language networks are spoken. The percolation simulation revealed that languages are likely to disappear in affluent countries, regardless of their economic disparities. One factor contributing to the likelihood of a future decline in languages in affluent countries is the influx of individuals to urban areas as a result of urbanization. With the movement from rural to urban areas, it is imperative to communicate and work in the language spoken in urban areas\cite{Mufwene2002}. This leads to a shift in language usage, wherein individuals from rural areas adopt the central language spoken in urban areas instead of their native language. It is probable that the advancement of technology will result in a significant decline in the number of languages in affluent countries.

\section*{Conclusion}
This study conducted the multilayer language-country bipartite network analysis using linguistic and economic data to quantitatively comprehend the characteristics of language endangerment, a global phenomenon in which historical, geographical, and economic factors are intricately intertwined. Based on the information pertaining to the countries in which a language is spoken, and two linguistic features, namely the existence of a writing system and the function within a country, a multilayer bipartite graph was generated. The analysis of the projected language and country networks revealed three characteristics of language endangerment. The initial characteristic was the remaining influence of colonialism on the language network, composed of officially used languages with their writing system. From the historical perspective, European languages, such as English, French, Spanish, Portuguese, Dutch, and Russian, were identified as central languages in the network. The subsequent characteristic was the geographical distribution of linguistic diversity. Unofficially used languages without a writing system created language networks in Asia, Africa, the Americas, and the Pacific, but not in Europe and multiple island countries, including Japan. This implied that the loss of linguistic diversity was caused by language policies based on nationalism. Furthermore, it was predicted that languages spoken in the Americas would be susceptible to the future language endangerment. The last characteristic was the relationship between the future extinction of languages and the wealth of countries. It was discovered that languages without a writing system were likely to disappear severely in countries with a high GDP per capita. These results show that network analysis has enabled quantitative analysis of language endangerment from historical, geographical, and economic perspectives.

\section*{Acknowledgments}
Authors acknowledge the financial support received from the Ripple Impact Fund (Grant Number: 2022-247584) for partial support to this work.

\nolinenumbers

%
%
%
\bibliographystyle{apa}
\bibliography{name.bib}

\end{document}